\documentclass[twocolumn,showpacs,amsmath,amssymb,aps,pre]{revtex4}   
\usepackage{graphicx}
\usepackage{dcolumn}
\usepackage{bm}
\usepackage{color}
\usepackage{amsmath}

\renewcommand{\vec}[1]{\mathbf{#1}}

\newif\ifgraph

\graphtrue             %

\begin{document}
\title{
Binary Mixtures in Linear Convection Arrays}

\author{Pulak K. Ghosh$^{1}$}
\author{Yunyun Li$^{2}$, Yuxin Zhou$^{2}$}
\author{Fabio Marchesoni$^{3}$}
\author{and Franco Nori$^{4,5}$}
\affiliation{$^{1}$ Department of Chemistry, Presidency University, Kolkata
700073, India}
 \affiliation{$^{2}$ Center for Phononics and Thermal Energy Science, Shanghai
 Key Laboratory of Special Artificial Microstructure Materials and Technology,
 School of Physics Science and Engineering, Tongji University, Shanghai 200092, China}
 \affiliation{$^{3}$ Dipartimento di Fisica, Universit\`{a} di Camerino, I-62032 Camerino, Italy}
 \affiliation{$^{4}$ Theoretical Quantum Physics Laboratory, RIKEN Cluster for Pioneering Research, Wako-shi, Saitama 351-0198, Japan}
 \affiliation{$^{5}$ Physics Department, University of Michigan, Ann Arbor, Michigan 48109-1040, USA}

\date{\today}

\begin{abstract}
We numerically investigated the dynamics of a mixture of finite-size active and passive particles in
a linear array of convection rolls. The interplay of advection and steric interactions
produces a number of interesting effects, like the stirring of a passive
colloidal fluid by a small fraction of slow active particles, or the separation of the mixture
active and passive colloidal fractions by increasing the motility of the active one,
which eventually clusters in stagnation areas along the array walls. These mechanisms are
quantitatively characterized by studying the dependence of the diffusion constants of the active and passive
particles on the parameters of the active mixture fraction.

\end{abstract}
\maketitle

\section{Introduction}

Brownian motion in convective laminar flows \cite{Moffatt} is a topic of
interdisciplinary research for its applications to transport
phenomena at large \cite{Chandra,Child2} and small scales
\cite{Kirby,PNAS}. For instance, under quite general conditions, advection
has been proven, both analytically \cite{Rosen,Soward,Shraiman,Pomeau} and
experimentally \cite{Gollub1,Gollub2,Saintillan1}, to enhance the diffusion of
passive colloidal particles along a linear array of convection rolls.
This is a combined effect 
of thermal noise and advection. Indeed, a noiseless particle would
keep circulatig inside the convection roll where it had been injected
 \cite{PoF}.

Active particles in a linear convection array behave differently.  An active
particle is modeled as a particle propelling itself with tunable constant
speed and direction changing because of collisions with obstacles and
other particles, or the torque associated with the convective flow, or
fluctuations (either of the suspension fluid or the self-propulsion
mechanism) \cite{Granick,Muller,Marchetti,Gompper}. A slow active particle
ends up trapped in a convection roll as long as the fluctuations affecting
its dynamics are negligible \cite{Neufeld}, whereas a fast active particle
tends to sojourn against the array walls, in the stagnation areas
separating the convection rolls. In the latter regime, the
particle diffusion along the array is dominated by the self-propulsion
mechanism \cite{PRR2,PCCP}.

So far, all studies on the advection-diffusion of colloidal suspension in convection
arrays were conducted in the low density regime, namely, they focused
on the diffusion of a single particle.  Of course, the dynamics of a {\em
suspension} of Brownian particles is in itself a well established topic due to its
applications in colloids and aerosols science \cite{Bird}.
Weakly attracting colloidal particles \cite{collrev2} are known to cluster
into a variety of continuously time-evolving structures, ranging from dimers
to crystals \cite{collrev1,collrev3}. Most remarkably, in the context of soft matter,
clustering of overdamped active particles in a stationary suspension fluid
can occur even in the absence of particle attraction, because the particles on
the cluster surface point mostly inwards \cite{Fily,Redner}. Under the same
conditions, steric interactions do not suffice to make passive particles
cluster; their distribution would remain uniform at all times.
However, despite the wide literature on colloidal suspensions, the impact of
collisions on particle transport in a convection array has been address only
recently \cite{SM22}. Extensive numerical simulations showed, for instance,
that particle-particle collisions allow convection cell crossings by
noiseless particles (active and passive, alike) and, therefore,
athermal diffusion in convection arrays and turbulent flows at large \cite{Tabeling}.

In this paper we investigate the collisional effects in a
binary mixture made of colloidal particles of the same size, one species
active and the other one passive. As mentioned above, single active and
passive particles advected in a linear convection array exhibit different
diffusion properties. On the other hand, steric collisions, due to the
particle finite size, allow the two mixture fractions to interact, so that the
question arises about the advection-diffusion of a binary mixture. Our numerical
simulations reveal a number of interesting new features, like the strong
stirring action exerted on a passive colloidal fluid by a small fraction of
slow active particles, or the separation of the mixture into two distinct
colloidal fluids for strong self-propulsion of the active
fraction, the passive fluid circulating inside the convection rolls and the active
one acccumulating in stagnation areas along the array walls.

To emphasize the interplay of advection and steric collisions, we neglected two
further aspects of the colloidal fluid dynamics, namely hydrodynamical and inertial
effects. Laminar flows around finite-size particles are likely to affect the advective
drag of the suspension fluid; therefore, the conclusions of the present paper should be
reconsidered at high packing fractions. Inertia in colloid advection
is also a well-established problem \cite{Stommel,Maxey1}, unavoidable in the case of
massive particles. However, neglecting both effects can be appropriate at 
low Reynolds numbers, when one deals with nano- and micro-particles \cite{Kirby}, 
as is often the case in soft matter systems.

The contents of this paper is organized as follows. In Sec. \ref{model} we
introduce the key ingredients of a two-dimensional (2D) model of advected
overdamped binary mixture consisting of active and passive hard disks of the
same size, namely, the fluid streaming function, the particle-particle
interaction function, and the self-propulsion mechanism of the active
fraction. Known results holding in the low-density (single particle) regime
are summarized in Sec. \ref{summary}. In Sec. \ref{1comp} we show that
particle collisions are responsible for spatial inhomogeneities
in pure passive and active suspensions confined into a linear convection
array. In Sec. \ref{2comp} we investigate the interaction between the active
and passive components of the mixture. The stirring effects of a small
fraction of slow active particles (Sec. \ref{stirr}) and the active-passive
separation caused by fast active particles (Sec. \ref{demix}) are discussed
in detail. In Sec. \ref{diffusion} the diffusion of the passive (Sec.
\ref{diffP}) and active particles (Sec. \ref{diffA}) are compared under both
stirring and demixing conditions. Finally, in Sec. \ref{conclusion} we draw a
few concluding remarks in view
of future work.

\begin{figure}[tp]
\centering \includegraphics[width=8.0cm]{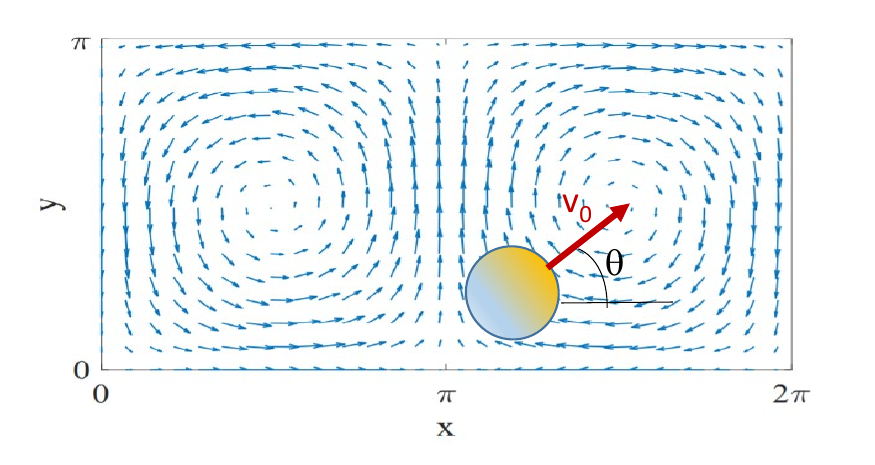}
\caption{Advection flow field, ${\vec v}_\psi$, in a linear convection array with the periodic stream function $\psi(x,y)$
of Eq. (\ref{psi}). The unit cell is delimited by horizontal
reflecting walls, $y=0$ and $y=L/2$, and periodic vertical boundaries, $x=0$ and $x=L$, with $L=2\pi$. The
self-propelling particle of  Eq. (\ref{LE}), is sketched for reader's convenience.}
\label{F1}
\end{figure}

\section{Model}\label{model}

Following Ref. \cite{Pomeau}, we model a 2D linear convection array as a
stationary laminar flow with stream function \cite{Stommel,Maxey1},
\begin{equation}
\label{psi}
\psi(x,y)= ({U_0L}/{2\pi})\sin({2\pi x}/{L})\sin({2\pi y}/{L}),
\end{equation}
periodic in the longitudinal $x$ direction and confined in the transverse
direction, $0\leq y\leq L/2$. The two parallel walls, $y=0$ and $y=L/2$, act
as dynamical reflecting boundaries. The unit cell of the stream function
consists of two counter-rotating convection rolls, see Fig.~\ref{F1}, with
advection velocity field ${\vec v}_\psi =(\partial_y, -\partial_x)\psi$. The
flow parameters (the spatial period, $L$, and the advection speed, $U_0$) can
be combined \cite{Pomeau} to define the advection diffusion scale,
$D_L=U_0L/2\pi$, and the roll vorticity, $\Omega_L=2\pi U_0/L$.

We simulated a colloidal suspension consisting of a mixture of $N$ hard disks
per unit cell, including a fraction $\eta$ of active disks. All disks, active
and passive alike, repel each other with potential function modeled by the
truncated-shifted Lennard-Jones function \cite{WCA},
\begin{eqnarray}\label{LJ}
V_{ij} &=& 4\epsilon\left[\left(\frac{\sigma}{r_{ij}}\right)^{12} -\left(\frac{\sigma}{r_{ij}}\right)^{6} \right], \;\; {\rm if}\;\;  r_{ij} \leq r_m \nonumber \\
  &=& 0 \;\; {\rm otherwise},
\end{eqnarray}
where $r_m=2^{1/6}\sigma$ and $\sigma$ represents the effective particle
``diameter''. Since our conclusions turn out to be rather insensitive to
the interaction strength parameter, $\epsilon=v_\epsilon \sigma^2/24$, we
set $v_\epsilon=1$ throughout the present numerical study \cite{SM22}.
This choice implies that particles interact only through
steric repulsion, i.e., all hydrodynamical interactions will be neglected.
Effects due to the actual geometry of the particles and the
self-propulsion mechanism can be encoded in the model parameters \cite{Gompper}.

As we are interested in colloidal mixtures at very low Reynolds numbers, the
dynamics of the $i$-th disk will be assumed to be overdamped (massless
particle approximation) and formulated in terms of two translational and
one rotational Langevin equation (LE),
\begin{eqnarray} \label{LE}
\dot {\vec r}_i&=& {\vec v}_{LJ, i}+ {\vec v}_{\psi, i} + {\vec v}_{0, i} +\sqrt{D_0}~{\bm \xi}_i(t) \\ \nonumber
\dot \theta_i &=& ({\alpha}/{2})~\nabla \times {\vec v}_{\psi, i} +\sqrt{D_\theta}~\xi_{\theta, i} (t),
\end{eqnarray}
where $i,j=1, \dots N$, and ${\vec r}_i=(x_i,y_i)$. Here, ${\vec v}_{\psi,
i}$ is the advection velocity introduced above, and ${\vec v}_{LJ, i}=
-\sum_{j=1}^N {\bm \nabla}_i V_{ij}$ is the collisional term due to pair
repulsion. For an active particle, the self-propulsion vector, ${\vec
v}_{0,i}=v_0(\cos \theta_i, \sin \theta_i)$, has constant modulus, $v_0$, and
is oriented at an angle $\theta_i$ with respect to the axis of the array
axis. Of course, for a passive particle, ${\vec v}_{0,i}=0$.  The flow shear
exerts a torque on the active particles proportional to the local fluid
vorticity, $\nabla \times {\vec v}_\psi$ \cite{Neufeld,PRR2}. For simplicity,
we adopt Fax\'en's second law, which, for an ideal no-stick spherical
particle, yields $\alpha=1$ \cite{Stark}. Of course, no shear torque is
exerted on a symmetric passive particles with $v_0=0$.

The translational thermal noises in the $x$ and $y$ directions, ${\bm
\xi}_i(t)=(\xi_{x,i}(t), \xi_{y,i}(t))$, for all particles, and the
rotational noise, $\xi_{\theta,i} (t)$, for the active particles only, are
stationary, independent, delta-correlated Gaussian noises, $\langle
\xi_{\mu,i}(t)\xi_{\nu,j}(0)\rangle = 2 \delta_{ij}\delta_{\mu,\nu}\delta
(t)$, with $\mu,\nu=x,y,\theta$. $D_0$ and $D_\theta$ are the respective
noise strengths, which for generality we assume to
be statistically unrelated~\cite{ourPRL}.

The LE (\ref{LE}) could be conveniently reformulated in dimensionless units
by rescaling $(x,y) \to (\tilde x, \tilde y)=(2\pi/L)(x,y)$ and $t \to \tilde
t= \Omega_L t$. After rescaling, the tunable model parameters would read
$\sigma \to (2\pi/L)\sigma$, $v_\epsilon \to v_\epsilon/U_0$, $v_0 \to
v_0/U_0$, $D_0 \to D_0/D_L$ and $D_\theta \to D_\theta/\Omega_L$.
Alternatively, upon setting $L=2\pi$ and $U_0=1$, our simulation results can
be regarded as expressed in dimensionless units, and then scaled back to
arbitrary dimensional units. The stochastic differential Eqs.~(\ref{LE}) for
$D_0>0$ and/or $D_\theta>0$, were numerically integrated by means of a standard
Milstein scheme \cite{Kloeden}. Particular caution was exerted in the
noiseless regime, $D_0=0$, because transients can grow exceedingly long, thus
affecting the computation of both the long-time spatial distribution of the
mixture in the array and the diffusion constants $D_{a,p}=\lim_{t\to \infty}
\langle [x(t) -x(0)]^2\rangle_{a,p} /2t$, respectively of the active and passive
particles, with the stochastic average, $\langle \dots \rangle_{a,p}$,
taken over the relevant mixture component.

\begin{figure}[tp]
\centering \includegraphics[width=7.5cm]{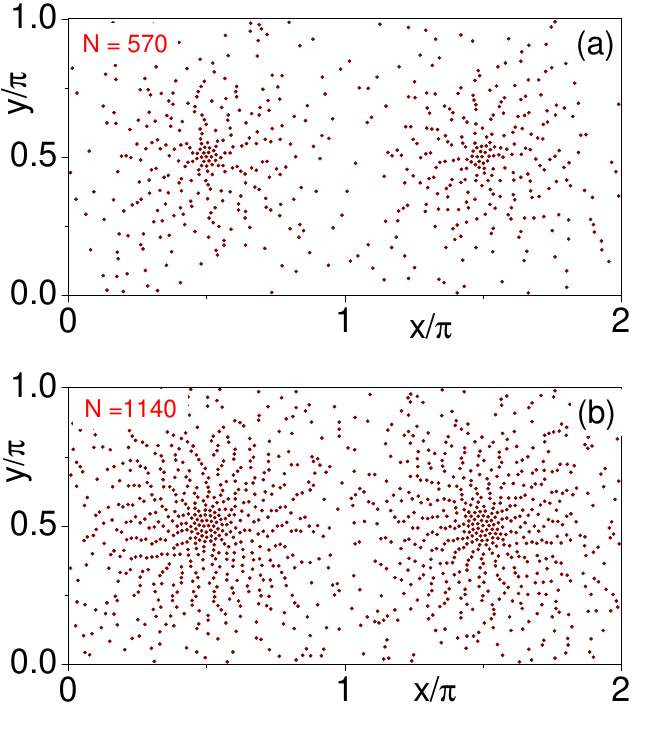}
\caption{Clustering of a suspension of noiseless passive particles, $D_0=v_0=0$,
after a running time $t=10^5$. At $t=0$ the suspension was uniformly distributed
with $N$ particles per unit cell of Fig. \ref{F1}. Other simulation
parameters are:  $\sigma=0.05$, $v_\epsilon=1$, $L=2\pi$ and $U_0=1$.}
\label{F2}
\end{figure}

\subsection{Earlier results} \label{summary}

Colloidal dynamics in convective flows has been studied (analytically and numerically)
mostly ignoring particle interactions XXX. A 2D fluid of
passive particles in a convection array was first investigated in Ref.
\cite{SM22} under the simplifying assumption that non-steric interactions
might be neglected. Particle collisions were shown to affect the colloidal
fluid dynamics well beyond the low density (single particle) approximation.

For this reason, we recall now the main properties of the diffusive dynamics
of a {\it single} overdamped colloidal particle in a convection array:

(i) In the absence of translational noise, $D_0=0$, an advected noiseless passive particle
gets trapped inside a single convection roll, where it keeps retracing the same closed
orbit, depending on its injection point \cite{Chandra,Stommel}.

(ii) In sharp contrast, for $D_0>0$ the spatial distribution of a
passive Brownian particle
along the convection rolls is uniform. Moreover, the translational noise activates particle hopping
between adjacent rolls with average hopping time inverse proportional to $D_0$ \cite{PoF}.

(iii) Accordingly, a {\em passive} Brownian particle undergoes normal
diffusion along the array with diffusion constant $D \sim \sqrt{D_L D_0}$,
where in our notation $D_0$ coincides with its free diffusion constant in the
suspension fluid at rest, and $D_L$ is the characteristic advection diffusion
constant defined above \cite{Rosen,Soward,Shraiman,Pomeau}. At high P\'eclet
numbers, $D_0\ll D_L$, advection enhances the spatial diffusion of the
particle by pushing it along the roll boundary layers
\cite{Gollub1,Gollub2,Saintillan1}.

(iv) Advection favors the confinement of noiseless {\em active} particles, as
well.   A noiseless active particle, $D_0=D_{\theta}=0$, ends up being trapped in a
convection roll, unless its speed is raised above a certain threshold value, $v_{\rm
th}$, proportional to the fluid advection speed, $U_0$ \cite{Neufeld}. For
$v_0 > v_{\rm th}$ the particle trajectory can be either bounded (but not
necessarily closed) or unbounded.  Spatial diffusion of an advected active
particle as a combined effect of rotational and thermal fluctuations has been
investigated in Refs. \cite{PRR2,PCCP} (see Sec. \ref{1comp} for
details relevant to the present report).

(v) The action of upward flows may suffice to counter the effect of a downward
transverse bias, for instance, of gravitational settling. Suppose a
suspended passive particle move under the action of the
external term, $-g {\bf \hat e}$ with ${\bf \hat e}=(0,1)$ and $g>0$, to be
inserted on the r.h.s. of the first LE (\ref{LE}). In the noiseless regime, $D_0=0$, The probability
the particle gets trapped in the cells of a 2D convection array (periodic in both
the $x$ and $y$ direction) depends on the ratio $g/U_0$ -- fully trapped for $g/U_0=0$
and no trapping for $g/U_0=1$ \cite{Stommel}\cite{Maxey1}. On the contrary, in the presence of translational noise,
its spatial distribution remains uniform. In
linear convection arrays the confining walls modify the spatial
distribution of an advected colloidal fluid,  active and passive
\cite{PREL,PCCP}, alike, as reported in the foregoing section.

Our numerical simulations confirmed that particle collisions play
an important role in the dynamics of an advected suspension, which led us to
reconsider some of the above results.

\begin{figure}[tp]
\centering \includegraphics[width=8.5cm]{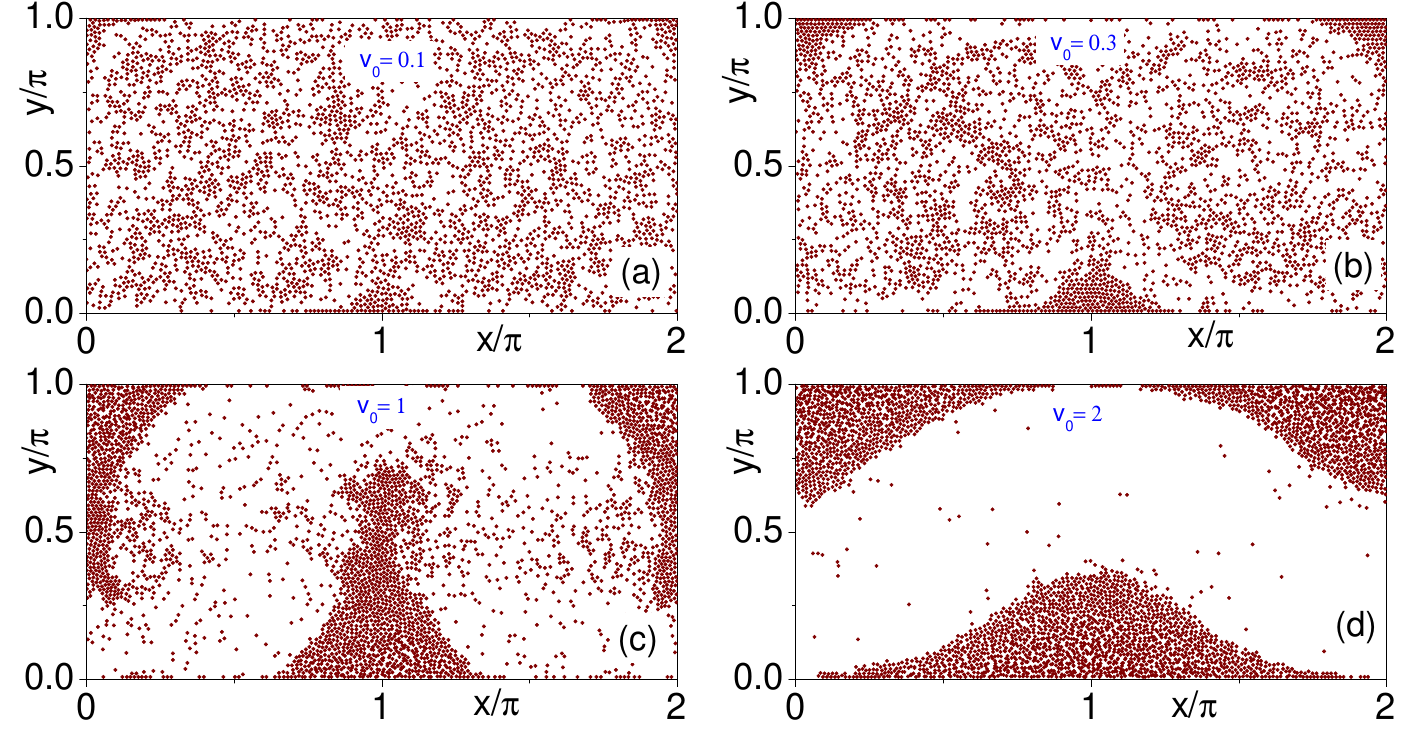}
\caption{Spatial distribution of a suspension of N=3420 noiseless active particles,
$D_0=\hat{}0$, with increasing speeds, $v_0$ (see legend),
after a running time $t=10^5$. At $t=0$ the suspension was uniformly distributed
with $N$ particles per unit cell of Fig. \ref{F1}. Other simulation
parameters are: $D_\theta=0.01$, $\sigma=0.05$, $v_\epsilon=1$, $L=2\pi$ and $U_0=1$.}
\label{F3}
\end{figure}

\section{One-component suspensions}\label{1comp}

In the context of soft matter, it has been
reported that clustering of overdamped active particles in a {\it stationary}
suspension fluid can occur even in the absence of particle attraction
\cite{Fily,Redner}. In sharp contrast, under the same conditions, finite-size passive
particles would not cluster (that is, not as an effect of steric interactions,
only); consistently with the single-particle picture, the suspension distribution
would remain uniform at all times.

\begin{figure}[bp] \centering \includegraphics[width=8.5cm]{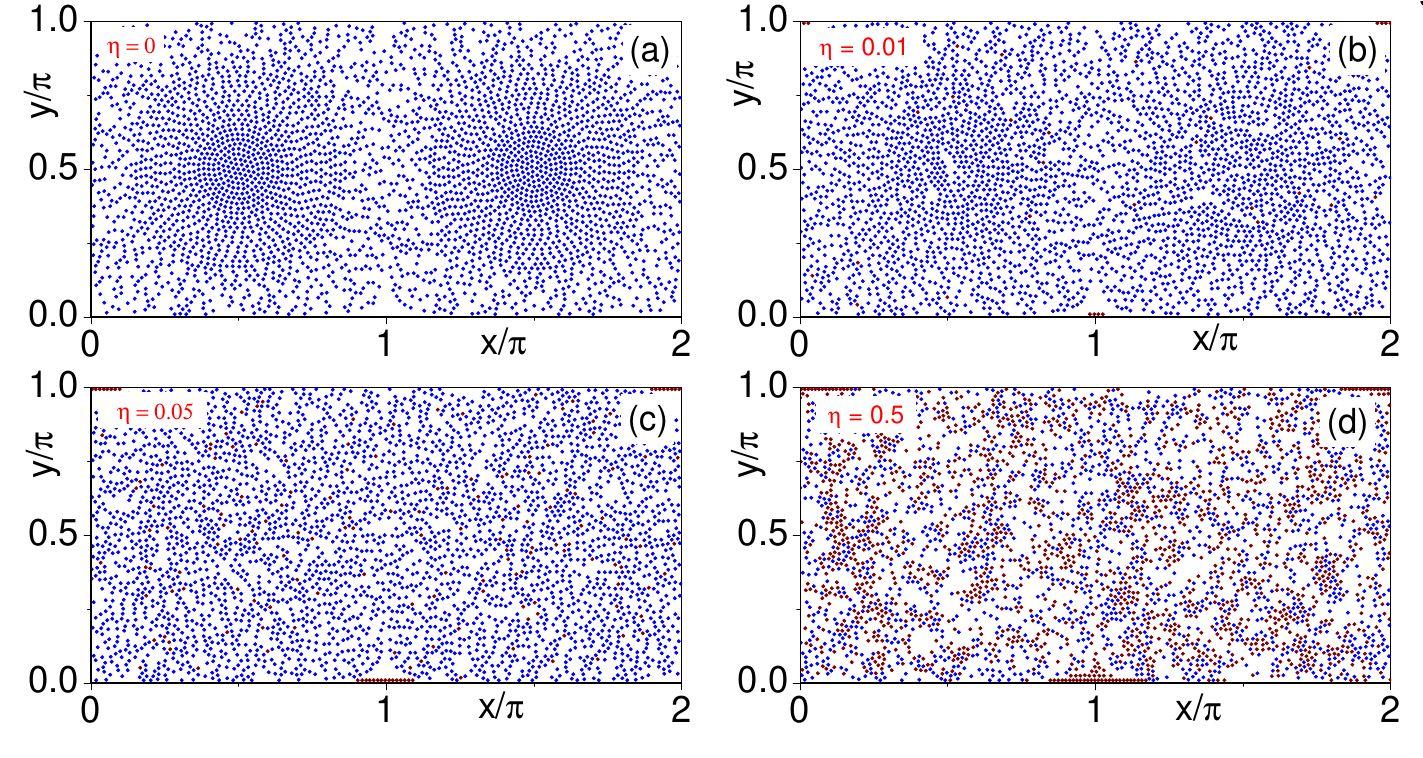}
\caption{Binary mixture of active (blue) and passive disks (brown) with
$D_0=0$, $D_\theta=0.01$ and $v_0=0.1$ at $t=10^{5}$. $N=3420$ is the total number of
disks, and $\eta$ (see legends) the active fraction.  At $t=0$ all $N$ disks
were randomly distributed in the $\psi(x,y)$ unit cell of Fig. \ref{F1}. All
other simulation parameters are as in Fig. \ref{F2}.} \label{F4}
\end{figure}

In a convective flow, a passive colloidal suspension behaves differently
\cite{SM22}. The spatial configuration of an initially uniform suspension of
hard disks advected in a linear convection array is displayed in Fig.
\ref{F2} [and \ref{F4}(a)] for increasing values of its density $2N/L^2$ (or
average packing fraction $\phi=\pi \sigma^2 N/2 L^2$). After a long running
time, say $t=10^5$, the particles have diffused toward the center of the
convection rolls, where they form stable rotating patterns, ranging from
micellar patterns at low $N$, to compact clusters at high $N$. Clusters
consist of a lattice core surrounded by ring-like formations and a dilute gas
of strongly advected particles in the vicinity of the cell boundaries. As
detailed in Ref. \cite{SM22}, the clustering process of a passive colloidal
fluid inside the convection rolls is governed by the inter-particle
collisions.

The clustering dynamics illustrated in Fig. \ref{F2} is independent of the
suspension preparation (not reported), but extremely sensitive to
translational noise with $D_0>0$. This result is consistent with the earlier
observation that, inside a convection roll, under periodic boundary conditions,
a single Brownian tracer approaches a uniform spatial distribution \cite{Neufeld,PoF}.

The behavior of a suspension of $N$ self-propelling disks in a linear convection
array is illustrated in Fig. \ref{F3}. The difference with the same density passive
suspension of Fig. \ref{F2} is apparent. As already reported for a single active JP
\cite{PRR3}, the disks, after reaching the channel walls, tend to slide along them
until they accumulate in the stable stagnation areas, namely at the base of the
ascending  and descending flows, respectively against the lower and upper walls.
Such accumulation areas are centered at $(x,y)=(L/2,0)$ and $(0,L/2)$, with $x$
given mod(L/2).

Contrary to the case of a single JP \cite{PRR3}, steric
interactions bring the disks in the stagnation areas to an almost complete
rest, the fraction of the floating disks decreasing with increasing $v_0$.
This behavior is apparent when self-propulsion wins over advection, that is
for $v_0/U_0>1$. Lowering $v_0$ for $v_0<U_0$, the disk accumulation along the channels
walls diminishes, as most of the disks are dragged along by the convective
flow. Contrary to the passive suspension of Fig. \ref{F2}, here the advected
disks seem to form small, short-lived clusters, which can be interpreted as a
residual effect of the athermal clustering effect reported in Ref. \cite{Fily}.
Advection disrupts the clustering action associated with self-propulsion;
clustering of slow active disks can only be achieved at much higher packing
fractions than in the absence of advection.

\begin{figure}[bp] \includegraphics[width=8.5cm]{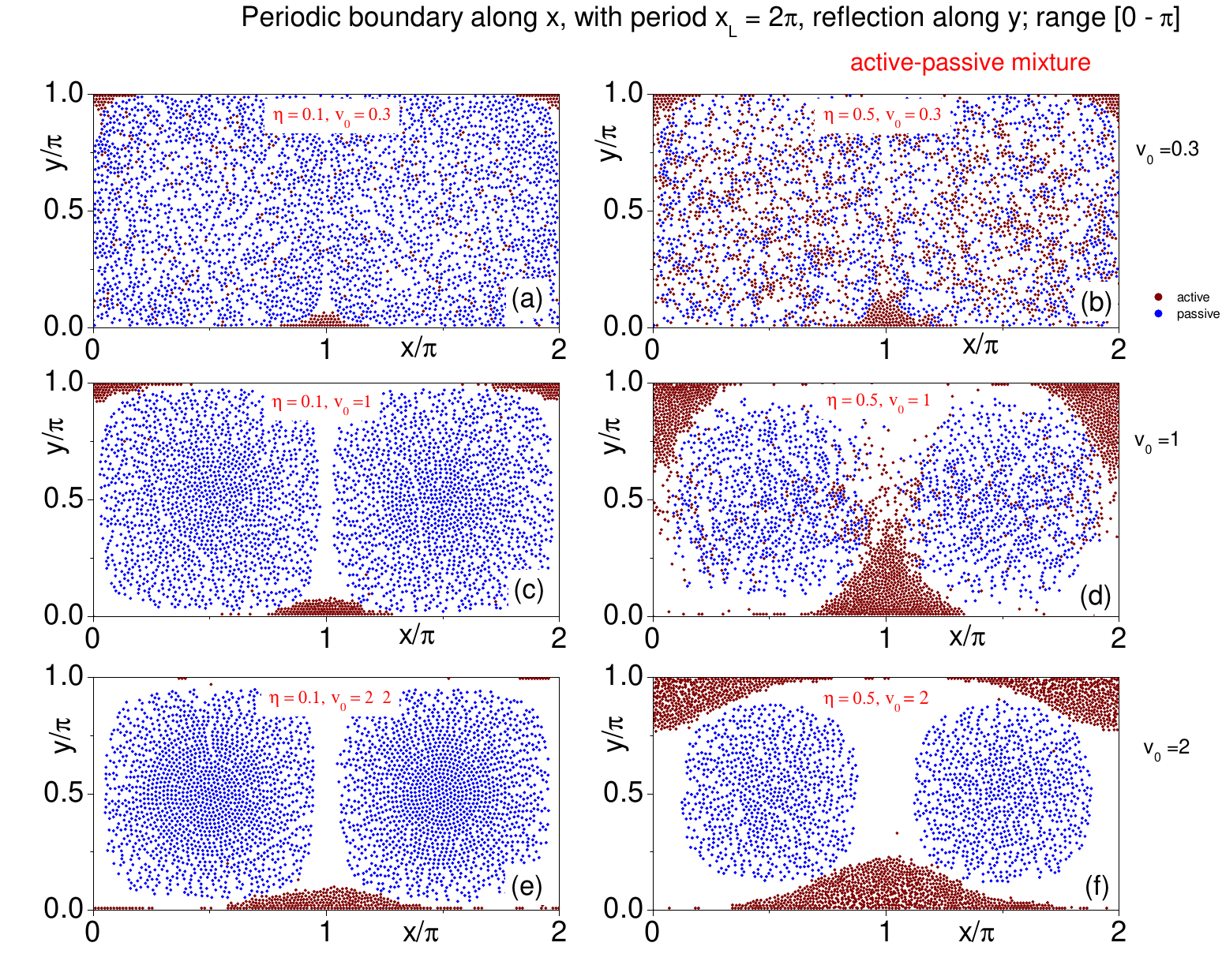}
\caption{Binary mixture of $N=3420$ active (blue) and passive disks (brown) with
$D_0=0$, $D_\theta=0.01$ and $v_0=0.3, 1.0$ and $2.0$ (from top to bottom): (a)-(c)
$\eta=0.1$ and (d)-(f) $\eta=0.5$.  At $t=0$ all $N$ disks were
randomly distributed in the $\psi(x,y)$ unit cell of Fig. \ref{F1}; all
snapshots were taken at $t=10^{5}$. All
other simulation parameters are as in Fig. \ref{F2}.} \label{F5}
\end{figure}
\section{Binary mixtures}\label{2comp}

Binary mixtures made of different active species have been the subject of
extensive numerical and analytical investigations for at least a decade now
\cite{mix1,mix2,mix3,mix4,mix5,mix6,mix7,stirr,Nanoscale}. The attention of most researchers
focused on the phenomenon known as motility induced phase separation (MIPS),
whereby the mixture components tends to separate as an effect of their
motility difference \cite{mix1,mix2}. Phase separation can occur on either
local or global scales, with the formation of one-species clusters of various
size. Clustering is initiated by the confining action exerted by the active
particles \cite{Fily,Redner}. However, MIPS is sensitive to other aspects of
the microscopic mixture dynamics, like hydrodynamical and phoretic effects
\cite{mix3}, chemical interactions \cite{mix4}, particle geometry
\cite{mix5}, and external confining fields of force \cite{mix6}, to mention
but a few. In the absence of these additional factors, MIPS emerges from a
delicate balance between steric interactions (the phase packing fractions)
and self-propulsion (their motility). Not surprisingly, this phenomenon has
been investigated for mixtures diffusing in equilibrium suspension fluids
and mostly in the absence of geometric constraints \cite{mix7}.

In the following we will study phase aggregation and segregation an
active-passive binary mixture advected in a linear convection array.

\subsection{Stirring} \label{stirr}

Consider a suspension of noiseless passive particles in the convection array of Eq.
(\ref{psi}). We know from Sec. \ref{1comp} that for not too small packing
fractions the particles tend to aggregate in regular patterns rotating around the center
of the convection rolls. Assume now to add a small amount
of slow active particles with the same geometry, and ask ourselves what is
their impact on the dynamics of the passive fluid.

We simulated numerically this situation for a mixture of $N=3420$ disks in
Fig. \ref{F4}, were the self-propelling disks have speed $v_0=0.1U_0$. An
active fraction as small as $\eta=0.01$, panel (b), suffices to destroy
the clustering pattern of the pure passive suspension, panel (a). As a result, for $\eta=0.05$ the passive
particles circulate inside the convection rolls with almost uniform spatial distribution.
Mixture stirring by active swimmers is a well-known mechanism \cite{stirr};
here, it can be invoked to mimic the effects of a tunable translational noise.

On further increasing the active fraction, $\eta$, the distributions of both
phases undergo micro-clustering. For $\eta=0.5$, panel (d), both species
aggregate in small, short-lived clusters. This effect is clearly due to the behavior of
the active component of the mixture reported in Sec. \ref{1comp}.
Note that the suspension of Fig. \ref{F3}(a) is the same as
the mixture in Fig. \ref{F4} but for $\eta=1$. Moreover, the emerging micro-clusters
are made of active and passive particles; no phase separation was detected on the
cluster scales, as long as the self-propulsion speed was kept sufficiently low.

\subsection{Demixing} \label{demix}

On increasing the speed, $v_0$, of the active disks, the scenario changes
dramatically, as illustrated in Fig. \ref{F5} for low and high values of the
active fraction, respectively, $\eta=0.1$ and $\eta=0.5$. The
micro-clustering phenomenon disappears for $v_0/U_0 \gtrsim 0.3$. Similarly
to the pure active suspension of Fig. \ref{F3}, for $v_0\gtrsim U_0$ self-propulsion wins over
advection, and the active disk pile up in the stagnation areas at
the base of the ascending (descending) flows against the lower (upper)
array walls. As a result, the passive disks, separated from the active
ones, diffuse back toward the center of the convection rolls, regrouping in
dynamical patterns, as first reported in Ref. \cite{SM22}, see also
Fig. \ref{F2}.

Phase demixing  grows more effective with increasing $v_0$, its onset being retarded
at large $\eta$. A large packing fraction of active disks
is harder to be contain in the stagnation areas separating the
convection rolls; hence the "fountain effect" in Figs. \ref{F3}(c) and
\ref{F5}(e). This explains why more active disks keep circulating inside the
convection rolls for $\eta =0.5$ than $\eta=0.1$, compare panels (b) and (e).
Demixing is complete only for $v_0\gg U_0$. However, even then, a small number
of active disks still happens to switch array wall. These are the disks stuck
in the outer layers of the aggregates grown along the walls. Owing to the
rotational fluctuation with time constant $D_\theta^{-1}$, they can self-propel
toward the center of the channel, thus crossing the passive
fluid circulating inside the convection rolls.
Traces of their passage are clearly visible as ``scars'' in the otherwise regular
patterns of the separated passive phase.

\section{Diffusion in a binary mixture}\label{diffusion}

We already mentioned that pair collisions in a colloidal fluid of appropriate
density suffice to make an individual particle cross the convection roll
separatrices even in the absence of thermal noise and self-propulsion. As
reported in Ref. \cite{SM22}, the particles of a noiseless passive suspension
may thus diffuse along a linear convection array as an effect
of steric collision alone (athermal diffusion).
On the other hand, advection  hinders the phenomenon of
athermal clustering that takes place in dense active suspensions under
no-flow conditions XXX \cite{Fily,Redner}.
Not surprisingly, in a binary mixture the interplay between active and passive phases,
affects the diffusion of both species.

\subsection{Passive particles} \label{diffP}

\begin{figure}[tp] \centering \includegraphics[width=8.5cm]{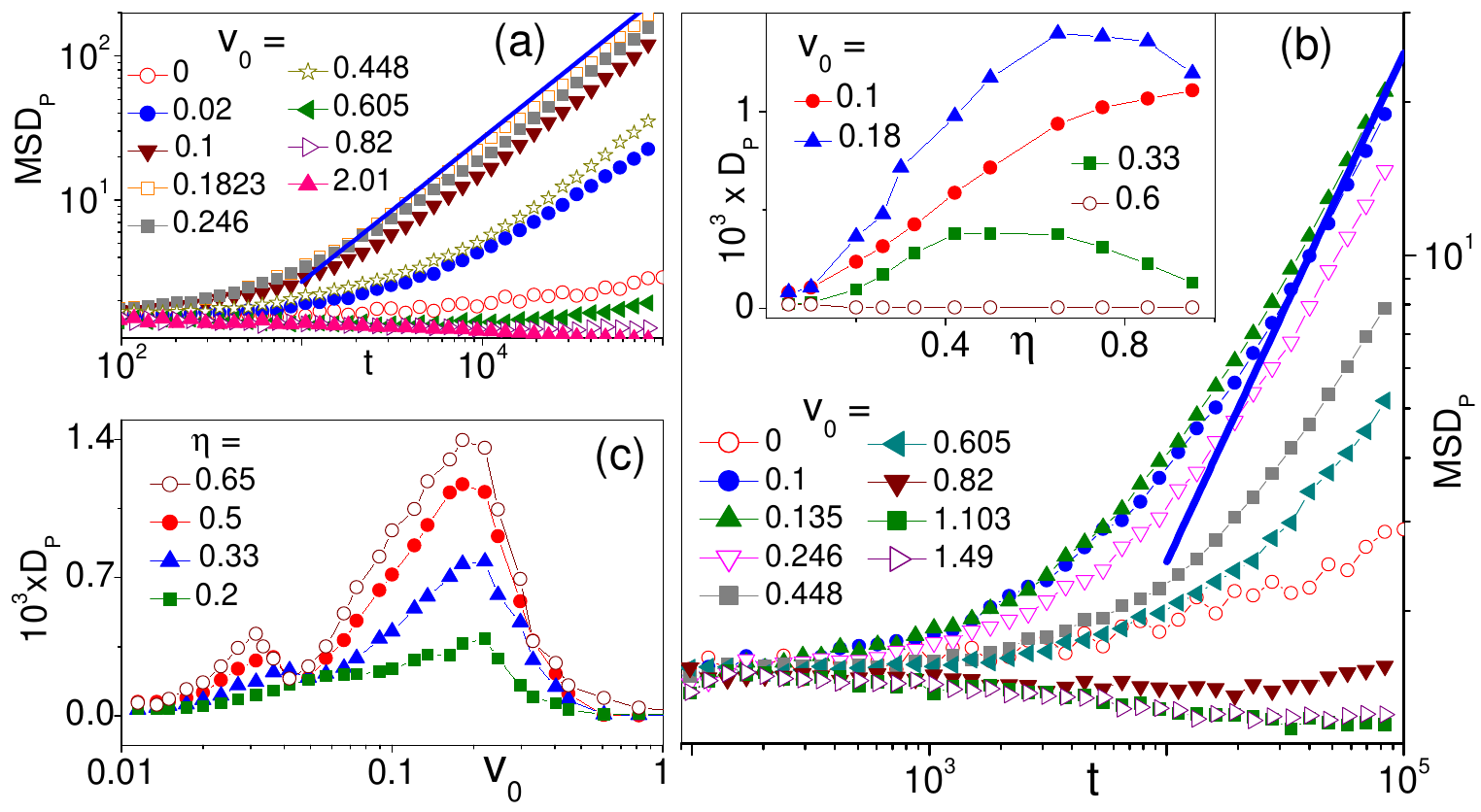}
\caption{Mean-square displacement, $\langle [x(t)-x(0)]^2\rangle_p$, vs. time, $t$,
of the passive disks in a binary mixture of $N=1140$ disks along the convection array of Fig.
\ref{F1} for different self-propulsion speeds, $v_0$, and  (a) $\eta=0.5$
and (b) $\eta=0.1$. The relevant diffusion constants, $D_p$, are plotted vs. $v_0$ for different $\eta$ in (c)
and vs. $\eta$ for different $v_0$ in the inset of (b).
The asymptotic normal diffusion power-law is represented by solid lines.
Other simulation parameters are: $D_0=0$, $D_\theta=0.01$, $\sigma=0.05$,
$v_\epsilon=1$, $L=2\pi$ and $U_0=1$.} \label{F6}
\end{figure}

The numerical characterization of the athermal diffusion of the passive phase
of a binary mixture of $N=1140$ disks is illustrated in Fig. \ref{F6}. The
mean-square displacement (MSD$_p$) of the passive disks grows asymptotically with
normal diffusion law, $\Delta x^2(t)=\langle [x(t)-x(0)]^2\rangle_p=2D_pt$, for
any choice of the tunable model parameters. The dependence of $D_p$ on the
self-propulsion speed of the active mixture fraction, $v_0$, is
plotted in panel (c) for different values of $\eta$, and, vice versa, its
dependence on $\eta$ for different values of $v_0$, in the inset of panel (b).

For $v_0=0$, the mixture is a passive fluid with $\eta=0$ and its diffusion
is purely athermal \cite{SM22}. In the opposite limit, $v_0\gg U_0$, the
active phase accumulates against the array walls, leaving the passive one
largely undisturbed. The passive particles then form regular patterns
that rotate around the center of the convection rolls; the area accessible
to them is restricted to the sinusoidal effective channel delimited by the lateral
aggregates of active particles. This suppresses the diffusion of
the passive mixture fraction almost completely, as apparent in Figs. \ref{F6}(a),(b) for $v_0\gtrsim U_0$.
Such an effect grows more prominent on increasing the active fraction, $\eta$.

Adding slow active disks to a passive suspension favours its diffusivity.
Accordingly, the curves of $D_p$ versus $v_0$, $D_p(v_0)$, exhibit two maxima, a smaller
one for $v_0/U_0\simeq 0.03$ and a higher one for $v_0/U_0\simeq
0.2$. We attribute such optimal diffusive regimes to two distinct mechanisms;
(i) the first $D_p$ peak is a signature of the stirring effect introduced in Sec.
\ref{stirr}, whereby the motility of the active disks destroys the dynamical
clusters of the passive disks localized at the center of the convection
rolls, thus favoring their diffusion along the array; (ii) the higher peak
is related with the formation of micro-clusters, with active and passive
disks clumping together into short-lived small-scale structures. The higher motility
of the active phase is thus transferred to the passive one, whose diffusivity
is thus enhanced \cite{Nanoscale}.

The maxima of the curves $D_p(v_0)$ have a distinct dependence on the
fraction, $\eta$, of the active mixture component, see inset of Fig.
\ref{F6}(b). The height of the smaller $D_p$ peak increases monotonically
with $\eta$, which suggests that the slow active disks stir the mixture
independently of each other, i.e., not through a collective action. The
position of the higher $D_p$ peak, $v_0\simeq 0.2$, is not sensitive to
$\eta$, but its height goes through a maximum for $\eta \simeq 0.7$. We
associate this diffusion peak with the breaking up of the circulation of the
active mixture fraction inside the convection rolls. Due to the steric
interactions, the active particles diffuse toward the outer boundaries of the
convection rolls and get advected parallel to the array walls
\cite{PRR2,PCCP}. In the process they drag along the passive particles,
whose diffusion is thus greatly enhanced. However, this is no longer the case
for larger values of $v_0$ (and $\eta$), when the active phase tends to
separate from the passive one, by aggregating against the walls.

\subsection{Active particles} \label{diffA}

\begin{figure}[tp] \centering \includegraphics[width=8cm]{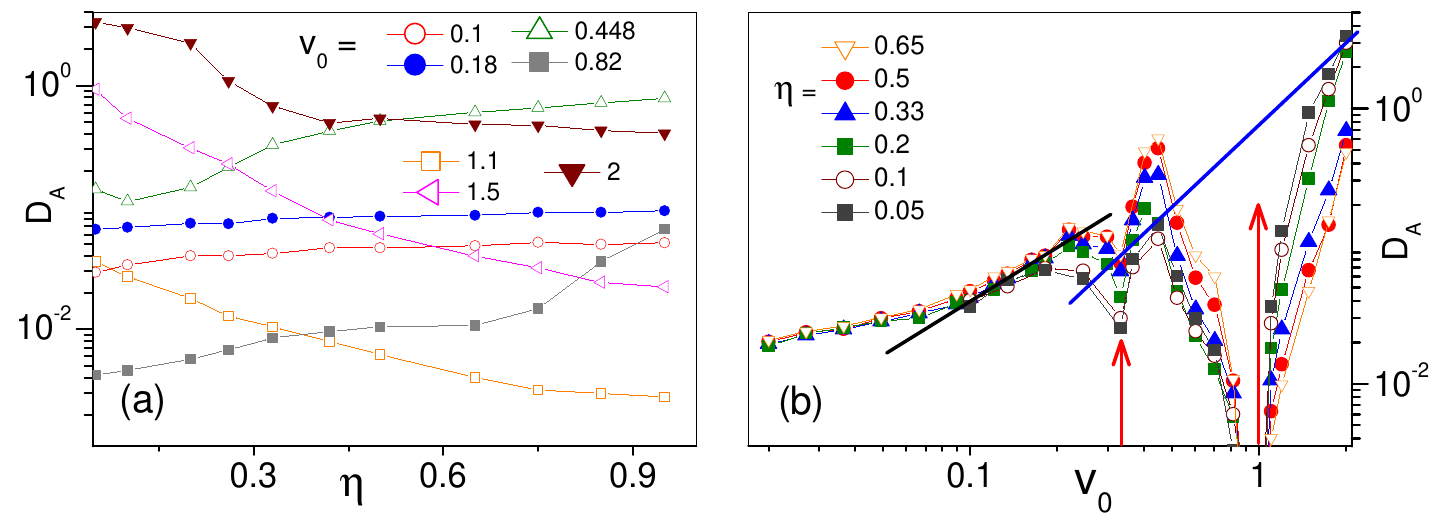}
\caption{Diffusion constant, $D_a$, of the active disks of the binary mixture of Fig. \ref{F6}:
(a) vs. $\eta$ for different $v_0$; (b) vs. $v_0$ for different $\eta$. The vertical arrows in
(b) correspond to the higher maxima of the curves $D_p(v_0)$ in Fig. \ref{F6}(c) (left) and the
single-particle depinning condition from the wall advective flow, $v_0=U_0$ (right).
Straight lines are linear and quadratic reference power-laws. \label{F7}}
\end{figure}

The different diffusion regimes of the passive component of the mixture have a clear-cut
signature in the $v_0$ dependence of the active diffusion constant, $D_a$, displayed in
Fig. \ref{F7} for different values of the active fraction, $\eta$. For $v_0\lesssim 0.2$,
$D_a$ is quite insensitive to the density of the active disks, panel (a), and grows
linearly with $v_0$, panel (b), as already observed for a single active particle in
Ref. \cite{PRR2}. The curves of panel (a) also show that increasing $\eta$ favors first
the depinning of the active disks from the convection rolls, signalled by an increase
of $D_a$, and then their aggregation in the appropriate stagnation areas along the channels
walls, with a consequent drop of $D_a$.

The relation between the $v_0$ dependence of $D_a$ and $D_p$ becomes apparent
when comparing Figs. \ref{F6}(c) and \ref{F7}(b). The growth of the diffusion
constant of a single active particle in a convection array is known to turn
from linear to quadratic \cite{PRR2}, as suggested in Fig. \ref{F7}(b). In
the presence of steric interactions, however two dips appear in the $D_a(v_0)$
curves: (i) one in correspondence with the higher maximum of $D_p(v_0)$,
which can be attributed to the "diffusivity transfer" \cite{Nanoscale}
between the active and passive mixture components; (ii) a deeper one for $v_0
\simeq U_0$, which marks the aggregation of the fast active disks at the
bases of the ascending (descending) flows along the lower (upper) array
walls. The active disks resume their typical motility with $D_a \propto
v_0^2$ only for much larger self-propulsion speeds, when the advection
effects grows negligible.

\section{Conclusions}\label{conclusion}

We have investigated the diffusion properties of a mixture of active and
passive colloidal particles of finite size advected in a linear convection
array. The combination of advection and steric collisions results in peculiar mixing
and demixing mechanisms, which depend on the motility and density of the active
fraction.

Besides its fundamental interest in the field of soft and biological matter,
this problem has practical implications, for instance, in medical
sciences. Let us consider a convection flow along a narrow channel, say, the vessel of
a biological organism. The particles advected by the flow can be either
passive drug complexes or active swimmers (synthetic or biological, alike), or a 
mixture thereof.
Based on the outcome of the present study, a
small fraction of active nano-particles can help control drug delivery by
preventing the passive suspension from clustering in the convection rolls.
Vice versa, an excess of more motile active swimmers tends to sediment in
stagnation areas against the channel walls, thus causing solid occlusions
that may hinder transport in the channel.

As the focus of this report was on the interplay of advective and collisional
dynamics, two important ingredients were ignored, namely, inertia and hydrodynamical
interactions. While this assumption may be justified at low Reynolds numbers,
appreciable inertial and hydrodynamical effects are expected in the case of large
and massive advected particles, irrespective of their motility. This question will
be addressed in a forthcoming publication.

\section*{Acknowledgements}
We thank RIKEN Hokusai for providing computational
resources. P.K.G. is supported by SERB Core Research Grant No. CRG/2021/007394.
Y.L. is supported by the NSF China under grants No. 11875201 and No.11935010.
F.N. is partially supported by Nippon Telegraph and Telephone Corporation (NTT) Research; the
Japan Science and Technology Agency (JST) via the
Quantum Leap Flagship Program (Q-LEAP) and the
Moonshot R\&D Grant No. JPMJMS2061; the Japan
Society for the Promotion of Science (JSPS) via the
Grants-in-Aid for Scientific Research (KAKENHI) Grant
No. JP20H00134; the Army Research Office (ARO)
(Grant No. W911NF-18-1-0358).

\end{document}